\documentclass[12pt]{article}
\usepackage{amssymb,amsmath,epsfig}
\allowdisplaybreaks

\begin{document}

\title{\bf Anisotropic Spherical Solutions in Rastall Gravity by Gravitational Decoupling}
\author{M. Sharif$^1$ \thanks{msharif.math@pu.edu.pk}~
and M. Sallah$^{1,2}$ \thanks{malick.sallah@utg.edu.gm} \\
$^1$ Department of Mathematics and Statistics, The University of Lahore\\
1-KM Defence Road Lahore-54000, Pakistan.\\
$^2$ Department of Mathematics, The University of The Gambia,\\
Serrekunda, P.O. Box 3530, The Gambia.}
\date{}
\maketitle

\begin{abstract}
In this paper, we extend the Finch-Skea isotropic ansatz
representing a self-gravitating interior to two anisotropic
spherical solutions within the context of Rastall gravity. For this
purpose, we use a newly developed technique, named as gravitational
decoupling approach through the minimal geometric deformation. The
junction conditions that provide the governing rules for the smooth
matching of the interior and exterior geometries at the hypersurface
are formulated with the outer geometry depicted by the Schwarzschild
spacetime. We check the physical viability of both solutions through
energy conditions for two fixed values of the Rastall parameter. The
behavior of the equation of state parameters, surface redshift and
compactness function are also investigated. Finally, we study the
stability of the resulting solutions through Herrera cracking
approach and the causality condition. It is concluded that the
chosen parametric values provide stable structure only for the
solution corresponding to the pressure-like constraint.
\end{abstract}

\section{Introduction}

The Rastall gravity theory proposed by Peter Rastall in 1972
\cite{1} has recently enjoyed a rebirth in popularity
\cite{7}-\cite{8}. This theory (a generalization of Einstein's
theory of general relativity) is based on the proposition that the
stress-energy tensor which exhibits null divergence in flat
spacetime is not always conserved in a curved spacetime geometry.
Rastall gravity deviates from general relativity because it
incorporates the Ricci scalar via the Rastall parameter. Despite
being manually introduced, this factor changes not only the field
equations but also the way material fields are coupled to the
gravitational interaction. It is obvious that the minimal coupling
principle does not hold true in this theory. However, this also
carries with it new and exciting insights that may help us to
comprehend a number of widely researched phenomena, including
cosmological problems, stellar systems, collapsed structures like
black holes, gravitational waves, etc. Rastall gravity is thus
equally competitive as other modified theories of gravity like
$f(\mathcal{R})$ and $f(\mathcal{R},T)$ theories, where
$\mathcal{R}$ and $T$ denote the Ricci scalar and trace of the
energy-momentum tensor, respectively.

It is worth mentioning that the $f(\mathcal{R},T)$ gravity
introduces matter and geometric terms, possessing the minimal as
well as non-minimal couplings. On the other hand, the Rastall theory
simply inserts geometric objects, specifically the Ricci scalar. To
evaluate, at least hypothetically, how well the results match the
widespread acceptance of general relativity, the consequences
produced by the additional terms have been intensively examined on
various fronts. Any perfect fluid solution of the Einstein field
equations is also a solution of the Rastall field equations, which
is a noteworthy aspect of the Rastall theory of gravity. With regard
to the black holes, both Rastall gravity and general relativity have
the same vacuum solution. The Rastall field equations, although
generalizing the field equations of general relativity, preserve the
theory's general coordinate transformation even though they lack an
associated Lagrangian density from which they may be derived.

In order to develop solutions that accurately represent the
gravitational behavior of such systems while taking into
consideration the non-minimal coupling, the interaction between
Rastall gravity and anisotropic spherical systems is investigated in
this study. We explore the search for anisotropic spherical
solutions through gravitational decoupling via minimal geometric
deformation (MGD) in this context. This technique has proven to be a
useful tool in addressing the problem of identifying interior
solutions for self-gravitating systems. The roots of this approach
can be found in \cite{9} within the setting of the brane-world
theory \cite{10}, which later extended to investigate new black hole
solutions \cite{11,12}. Ovalle and Linares \cite{13} developed an
exact interior solution for isotropic spherically symmetric compact
distributions, which is effectively a brane-world adaptation of
Tolman's solution. Casadio et al. \cite{14} discussed a unique
external solution for spherical self-gravitating systems having a
naked singularity at the Schwarzschild radius to adjust temporal and
radial metric functions. Ovalle \cite{15} further built anisotropic
solutions from ideal fluid configuration with spherical symmetry,
using the same technique. Ovalle and collaborators \cite{16}
expanded isotropic interior solutions to take the impacts of
anisotropy into account. The gravitational decoupling technique
comes in two folds, namely MGD and the extended geometric
deformation (EGD). The former (MGD) deforms only the radial
component of the metric while the latter (EGD) deforms both the
temporal and radial metric components. It is also worthy of mention
that these deformations are introduced via some appropriate linear
transformations of the spacetime metric components.

Many physical events indicate that pressure anisotropy is a key
factor to check how stellar bodies evolve. By taking into account a
certain type of anisotropy, some researchers \cite{17} were able to
get precise solutions, demonstrating that spherical stars may
sustain positive and finite pressures and densities while also
offering insights on practical astrophysical objects. In their study
of anisotropic self-gravitating spheres, Gleiser and Dev \cite{18}
showed that anisotropy can support stars with a particular
compactness $\frac{M}{2R}=\frac{2}{9}$, ($M$ is the mass of the star
and $R$ denotes the radius) and came to the conclusion that stable
configurations exist for particular adiabatic index in comparison to
isotropic fluids. Sharma and Maharaj \cite{19} made a substantial
progress in the modeling of compact stars by getting accurate
solutions for spherically symmetric anisotropic matter distributions
meeting a linear equation of state (EoS).

Herrera \cite{20} introduced the concepts of ``cracking'' and
``overturning'' to study the behavior of isotropic and anisotropic
structures after disturbances, as part of research into the
stability of self-gravitating models. His results showed that while
anisotropic fluid distributions crack, ideal fluid distributions
remain stable. In order to investigate anisotropic spherical
structures, Abreu et al. \cite{21} modified the idea of cracking by
integrating sound speed. They came to the conclusion that the system
becomes unstable when the square of the tangential sound speed
exceeds that of the radial sound speed. The physical relevance of
the MGD-decoupling method is highlighted by virtue of its
applicability in a variety of scenarios such as Einstein-Maxwell
systems \cite{22}, Einstein-Klein-Gordon systems
\cite{23}-\cite{26}, higher derivative gravity \cite{27}-\cite{29},
$f(R)$ theory \cite{30}-\cite{36}, Horava-aether gravity
\cite{37,38}, and polytropic spheres \cite{39}-\cite{41}, among
others. One of the simplest practical uses of MGD-decoupling is to
maintain the physical viability of existing isotropic interior
solutions for spherically symmetric self-gravitating systems in the
anisotropic domain, as highlighted in \cite{42}. The MGD technique
is an effective tool for obtaining anisotropic solutions in complex
gravitational systems while maintaining physical realism.

Henceforth, this paper proceeds with the following structural
organization. Section \textbf{2} deals with analysis of the Rastall
field equations for a static spherically symmetric matter
distribution and identifies effective parameters. In section
\textbf{3}, we use the MGD technique to split the Rastall field
equations into two simpler sets. The junction conditions that govern
the matching of the interior and exterior spacetimes are also
investigated. In section \textbf{4}, we obtain two solutions for
anisotropic spherical source, by extending a known perfect fluid
ansatz. Furthermore, physical characteristics ranging from viability
to stability are investigated for our obtained solutions. Finally, a
summary of our results and some concluding remarks are discussed in
section \textbf{5}.

\section{Rastall Theory of Gravity}

The Rastall gravity theory \cite{1} spurs from the refutal of the
fundamental assumption that the stress-energy tensor freely diverges
in a curved spacetime. The Rastall field equations, given by
\begin{equation}\label{1}
\mathcal{R}_{\tau\upsilon}-\frac{1}{2}\mathcal{R}g_{\tau\upsilon}=\kappa
(T_{\tau\upsilon}-\lambda\mathcal{R}g_{\tau\upsilon}),
\end{equation}
are consistent with the assumption that
\begin{equation}\label{2}
\nabla_\upsilon\,T^{\tau\upsilon}=\lambda
g^{\tau\upsilon}\nabla_\upsilon \mathcal{R},
\end{equation}
and reduce to Einstein's field equations in the event $\lambda=0$.
In the above equations, $\kappa$ indicates the coupling constant and
$\lambda$ is the Rastall parameter which creates the diversion from
general relativity and through which the Ricci scalar is
non-minimally coupled into the theory. The non-conservation of the
stress-energy tensor \eqref{2} as proposed by Rastall, induces a
non-minimal coupling between matter and geometry. By defining
\begin{equation}\label{3}
\bar{T}_{\tau\upsilon}=T_{\tau\upsilon}-\lambda\mathcal{R}g_{\tau\upsilon}\,,
\end{equation}
we can rewrite the field equations \eqref{1} as
\begin{equation}\label{4}
G_{\tau\upsilon}=\kappa\bar{T}_{\tau\upsilon}\,.
\end{equation}
This shows that the original Rastall field equations can always be
restructured to remold the Einstein's field equations, hence
regaining the standard result
$\nabla_\upsilon\bar{T}^{\tau\upsilon}=0$. This restructuring can
also be performed in other modified gravity theories such as
$f(\mathcal{R})$, $f(\mathcal{R},T)$ theories among others,
irrespective of the conservation of the stress-energy tensor.

Upon contracting the field equations \eqref{1}, we can write the
Ricci scalar as
\begin{equation}\label{5}
\mathcal{R}=\frac{\kappa T}{4\lambda\,\kappa-1}\,,
\end{equation}
which can, in turn, be used to rewrite the effective stress-energy
tensor \eqref{3} as
\begin{equation}\label{6}
\bar{T}_{\tau\upsilon}=T_{\tau\upsilon}-
\frac{\epsilon\,T}{4\,\epsilon -1}\,g_{\tau\upsilon}\,,
\end{equation}
where $\epsilon=\lambda\,\kappa$. For simplicity, we take $\kappa=1$
so that $\epsilon=\lambda$. At this point, it is clear that
$\lambda=\frac{1}{4}$ depicts a non-realistic scenario and must
therefore be avoided. Here, $T_{\tau\upsilon}$ is considered as a
perfect fluid matter configuration given by
\begin{equation}\label{7}
T_{\tau\upsilon}=(\rho+P)u_\tau u_\upsilon - P\,g_{\tau\upsilon}\,,
\end{equation}
where $u^\tau=\sqrt{g^{00}}\delta^\tau_0$ is the fluid 4-velocity
while $\rho$ and $P$ represent the energy density and isotropic
pressure, respectively. The components of the effective
stress-energy tensor \eqref{6} are thus obtained as
\begin{align}\label{8}
\bar{T}_{00}&=g_{00}\bigg(\frac{3\lambda(\rho+P)-\rho}{4\lambda-1}\bigg),\\\label{9}
\bar{T}_{11}&=-g_{11}\bigg(\frac{\lambda(\rho+P)-P}{4\lambda-1}\bigg),\\\label{10}
\bar{T}_{22}&=-g_{22}\bigg(\frac{\lambda(\rho+P)-P}{4\lambda-1}\bigg).
\end{align}
We now shift our attention to the field equations for multiple
matter sources, given by
\begin{equation}\label{11}
R_{\tau\upsilon}-\frac{1}{2}R\,g_{\tau\upsilon}=T_{\tau\upsilon}^{(tot)}\,,
\end{equation}
with
\begin{equation}\label{12}
T_{\tau\upsilon}^{(tot)}=\bar{T}_{\tau\upsilon}+\delta\,\Theta_{\tau\upsilon}.
\end{equation}
Here, $\bar{T}_{\tau\upsilon}$ is the usual matter sector for the
Rastall gravity given by Eq.\eqref{6} and the term
$\Theta_{\tau\upsilon}$ is an additional source gravitationally
coupled to the seed source through the constant $\delta$, that may
generate anisotropy in self-gravitating fields. New fields such as
scalar, tensor and vector fields may well be contained in the source
$\Theta_{\tau\upsilon}$. The addition of an extra source to a static
spherically symmetric gravitational source (usually referred to as
the seed source) is the foundation of the gravitational decoupling
procedure \cite{9,15}. Through this procedure, we can extend the
domain of known isotropic solutions (usually specified by the seed
source) to the domain of anisotropic configurations. The extra
source is thus responsible for including the effects of anisotropy
in the given configuration. We have thus employed this technique to
search for anisotropic spherical solutions, hence the justification
for Eq.\eqref{12}. By virtue of its definition, the total
energy-momentum tensor Eq.\eqref{12} must now satisfy the
conservation equation given by
\begin{equation}\label{13}
T^{\tau\,(tot)}_{\upsilon\,;\,\tau}=0\,.
\end{equation}

For the purpose of describing our interior geometry, we shall
consider a static spherically symmetric spacetime in
Schwarzschild-like coordinates as
\begin{equation}\label{14}
ds^2_{-}=e^{\alpha(r)}dt^2-e^{\beta(r)}dr^2-r^2(d\theta^2+\sin^2\theta
d\phi^2),
\end{equation}
where the areal radius $r$ ranges from the stars center $(r=0)$ to
an arbitrary point $(r=R)$ on the surface of the star. The
corresponding Rastall field equations turn out to
\begin{eqnarray}\label{15}
e^{-\beta}\bigg(\frac{\alpha^\prime}{r}-\frac{1}{r^2}\bigg)
+\frac{1}{r^2}=\frac{3\lambda(\rho+P)-\rho}{4\lambda-1}+
\delta\,\Theta^0_0\,,\\
\label{16}
e^{-\beta}\bigg(\frac{\alpha^\prime}{r}+\frac{1}{r^2}\bigg)
-\frac{1}{r^2}=\frac{\lambda(\rho+P)-P}{4\lambda-1} -
\delta\,\Theta^1_1\,, \\\label{17}
e^{-\beta}\bigg(\frac{\alpha^{\prime\prime}}{2}+\frac{\alpha^{\prime^2}}{4}
-\frac{\alpha^\prime \beta^\prime}{4}+\frac{\alpha^\prime -
\beta^\prime}{2r}\bigg)=\frac{\lambda(\rho+P)-P}{4\lambda-1} -
\delta\,\Theta^2_2\,.
\end{eqnarray}
With respect to the system \eqref{15} - \eqref{17}, the conservation
equation in \eqref{13} now reads
\begin{equation}\label{18}
\bar{P}^\prime(r)+\frac{\alpha^\prime(r)}{2}\,(\bar{\rho}+\bar{P}) +
\frac{2\delta}{r}\,(\Theta^2_2-\Theta^1_1)+\,\frac{\delta\alpha^\prime(r)}{2}
(\Theta^0_0-\Theta^1_1)-\delta\bigg(\Theta^1_1(r)\bigg)^\prime=0\,,
\end{equation}
where $\bar{\rho}=\frac{3\lambda(\rho+P)-\rho}{4\lambda-1}$ and
$\bar{P}=\frac{\lambda(\rho+P)-P}{4\lambda-1}.$ The Rastall field
equations \eqref{15}-\eqref{17} constitute a system of three
non-linear differential equations with seven unknowns namely two
physical variables $\rho(r)$ and $P(r)$, two geometric functions
$\alpha(r)$ and $\beta(r)$, and the functions
$\Theta^0_0\,,~\Theta^1_1\,,\Theta^2_2$ which constitute three
independent components of $~\Theta_{\tau\upsilon}$. Additionally,
the prime notation denotes the derivative with respect to the radial
coordinate, $r$. From this system, we identify three effective
matter components given by
\begin{align}\label{19}
\rho^{eff}=\rho+\delta\Theta^0_0,\quad
P_r^{eff}=P-\delta\Theta^1_1,\quad P_t^{eff}=P-\delta\Theta^2_2.
\end{align}
These definitions of the effective parameters indicate that the
source $\Theta_{\tau\upsilon}$ can instigate an anisotropy within
the stellar distribution given by
\begin{equation}\label{20}
\Delta=P_t^{eff}(r)-P_r^{eff}(r)=\delta(\Theta^1_1-\Theta^2_2).
\end{equation}
We now proceed to the next section where we shall explore the MGD
technique in a bid to demystify the field equations
\eqref{15}-\eqref{17}.

\section{Gravitational Decoupling Technique}

Using this approach, the field equations will split into two sets:
the first one (with $\delta=0$) given by the standard Rastall
equations for a perfect fluid while the second set will contain the
extra source $\Theta_{\tau\upsilon}$. To this effect, we consider a
perfect fluid solution
$\left\lbrace\eta\,,\sigma\,,\rho\,,P\right\rbrace$ of the field
equations \eqref{15}-\eqref{17}, where $\eta$ and $\sigma$ denote
the corresponding metric functions. Therefore, the metric in
Eq.\eqref{14} now reads
\begin{equation}\label{21}
ds^2=e^{\eta(r)}\,dt^2-\frac{1}{\sigma(r)}\,dr^2-r^2(d\theta^2+\sin^2\theta\,d\phi^2)\,,
\end{equation}
with
\begin{equation}\label{22}
\sigma(r)=1-\frac{2m(r)}{r},
\end{equation}
where $m$ represents the Misner-Sharp mass function. To inculcate
the effects of the source $\Theta_{\tau\upsilon}$ on the perfect
fluid solution, we consider the following minimal geometric
deformation
\begin{equation}\label{23}
\eta(r)\mapsto\alpha(r)=\eta(r),\quad\sigma(r)\mapsto
e^{-\beta(r)}=\sigma(r)+\delta h^\ast(r),
\end{equation}
where $h^\ast$ is the deformation endured by the radial component of
the metric function. Thus it is seen that only the radial metric
component in Eq.\eqref{21} is deformed whilst the temporal component
remains unaltered. Substituting the minimally deformed radial
coefficient from Eq.\eqref{23} into the field equations
\eqref{15}-\eqref{17}, the system splits into the following two sets
as foretold in the beginning of this section.

The first set reads
\begin{align}\label{24}
\frac{1}{r^2}-\frac{\sigma}{r^2}-\frac{\sigma^\prime}{r}&=
\frac{3\lambda(\rho+P)-\rho}{4\lambda-1},\\\label{25}
\sigma\bigg(\frac{\eta^\prime}{r}+\frac{1}{r^2}\bigg)
-\frac{1}{r^2}&=\frac{\lambda(\rho+P)-P}{4\lambda-1},\\\label{26}
\sigma\bigg(\frac{\eta^{\prime\prime}}{2}+\frac{\eta^{\prime^2}}{4}
+\frac{\eta^\prime}{2r}\bigg)+\sigma^\prime\bigg(\frac{\eta^\prime}{4}
+\frac{1}{2r}\bigg)&=\frac{\lambda(\rho+P)-P}{4\lambda-1},
\end{align}
with the associated conservation equation given as
\begin{equation}\label{27}
\bar{P}^\prime(r)+\frac{\eta^\prime(r)}{2}\,(\bar{\rho}+\bar{P})=0.
\end{equation}
Equations \eqref{24} and \eqref{25} can be solved simultaneously in
order that the quantities $\rho$ and P might be explicitly expressed
as functions of the metric potentials only. Thus we have
\begin{eqnarray}\label{28}
\rho&=&\frac{1}{r^2}-\frac{\sigma^\prime}{r}-\frac{\sigma}{r^2}
-\lambda\bigg[\frac{4}{r^2}-\frac{\sigma^\prime}{r}
-\sigma\bigg(\frac{3\eta^\prime}{r}+\frac{4}{r^2}\bigg)\bigg]\,,
\\\label{29}
P&=&-\frac{1}{r^2}+\sigma\bigg(\frac{1}{r^2}+\frac{\eta^\prime}{r}\bigg)
+\lambda\bigg[\frac{4}{r^2}-\frac{\sigma^\prime}{r}
-\sigma\bigg(\frac{3\eta^\prime}{r}+\frac{4}{r^2}\bigg)\bigg]\,.
\end{eqnarray}
The second set of equations (corresponding to the source
$\Theta_{\tau\upsilon}$) reads
\begin{eqnarray}\label{30}
\Theta_0^0&=&-\frac{h^{\ast^\prime}}{r}-\frac{h^\ast}{r^2}\,
\\\label{31}
\Theta^1_1&=&-h^\ast\bigg(\frac{\eta^\prime}{r}+\frac{1}{r^2}\bigg)\,,
\\\label{32}
\Theta^2_2&=&-h^\ast\bigg(\frac{\eta^{\prime\prime}}{2}
+\frac{\eta^{\prime^2}}{4}+\frac{\eta^\prime}{2r}
\bigg)-h^{\ast^\prime}\bigg(\frac{\eta^\prime}{4}+\frac{1}{2r}\bigg),
\end{eqnarray}
and satisfies the conservation equation given by
\begin{equation}\label{33}
\frac{2}{r}(\Theta^2_2-\Theta^1_1)+\frac{\eta^\prime(r)}{2}(\Theta^0_0-\Theta^1_1)
-\bigg(\Theta^1_1(r)\bigg)^\prime=0.
\end{equation}
It can be observed that the system Eqs.\eqref{30}-\eqref{32} above
comprises three equations in the four unknowns
$\big(\theta_0^0,\theta^1_1,\theta_2^2,h^\ast\big)$. It is worthy of
noting that $\eta$ is not considered an unknown in this system as it
will be evaluated from the system Eqs.\eqref{24}-\eqref{26}, the
first system after the decoupling process. It thus suffices to
impose a single constraint to evaluate the anisotropic system
Eqs.\eqref{30}-\eqref{32}. Consequently (in section 4) two
constraints are employed on the extra source $\Theta_{\tau\nu}$, and
in each case a solution is obtained.

We now shift our attention to the junction conditions which provide
the governing rules for the smooth matching of the interior and
exterior space-time geometries at the surface of the star (where
$r=R$). Our interior spacetime geometry is given by the deformed
metric
\begin{equation}\label{34}
ds^2_-=e^{\eta(r)}dt^2-\left(1-\frac{2m(r)}{r}+\delta
h^\ast(r)\right)^{-1}dr^2 -r^2(d\theta^2+\sin^2\theta d\phi^2),
\end{equation}
which is to be matched with the general outer metric given by
\begin{equation}\nonumber
ds^2_+=e^{\eta(r)}dt^2-e^{\beta(r)}dr^2-r^2(d\theta^2+\sin^2\theta
d\phi^2).
\end{equation}
Hence, the continuity of the first fundamental form
$\left([ds^2]_{\Sigma}=0\right)$ of junction conditions at the
hypersurface $\Sigma$ yields
\begin{equation}\label{35}
\eta(R)_-=\eta(R)_+,
\end{equation}
and
\begin{equation}\label{36}
1-\frac{2\,M_0}{R}+\delta\,h^\ast_R=e^{-\beta(R)_+},
\end{equation}
where $M_0=m(R)$ and $h^\ast_R$ is the deformation at the surface of
the star. Similarly, the continuity of the second fundamental form
($[T_{\tau\upsilon}S^\upsilon]_{\Sigma}=0,~S^\upsilon$ denotes a
unit 4-vector) gives
\begin{equation}\label{37}
P(R)-\delta\left(\Theta^1_1(R)\right)_-=-\delta\left(\Theta^1_1(R)\right)_+.
\end{equation}
Substituting Eq.\eqref{31} for the interior geometry in \eqref{37}
yields
\begin{equation}\label{38}
P(R)+\delta\,h^\ast(R)\bigg(\frac{\eta^\prime(R)}{R}+\frac{1}{R^2}\bigg)
=-\delta\left(\Theta^1_1(R)\right)_+.
\end{equation}
Using Eq.\eqref{31} for the outer geometry in \eqref{38}, we obtain
\begin{equation}\label{39}
P(R)+\delta\,h^\ast(R)\bigg(\frac{\eta^\prime(R)}{R}+\frac{1}{R^2}\bigg)
=\delta\,b^\ast(R)\bigg[\frac{1}{R^2}+\frac{2\mathcal{M}}
{R^3\big(1-\frac{2\mathcal{M}}{R}\big)}\bigg],
\end{equation}
where $\mathcal{M}$ is the mass in the exterior region and
$b^\ast(R)$ is the minimal geometric deformation inflicted on the
outer Schwarzschild solution by the source $\Theta_{\tau\upsilon}$,
as shown below
\begin{equation}\label{40}
ds^2_+=\left(1-\frac{2\mathcal{M}}{r}\right)dt^2 -
\left(1-\frac{2\mathcal{M}}{r}+\delta b^\ast(r)\right)^{-1}dr^2-r^2
d\Omega^2.
\end{equation}
Thus in essence, the extra energy-momentum tensor
($\Theta_{\tau\upsilon}$) contributes from both inside and outside
the interior distribution of matter, as have been portrayed by
Eqs.\eqref{38} and \eqref{39}, respectively. Equations \eqref{35},
\eqref{36} and \eqref{39} are the necessary and sufficient
conditions for the smooth matching of the deformed interior metric
\eqref{34} to the deformed spherically symmetric vacuum
Schwarzschild metric \eqref{40}.

\section{Anisotropic Spherical Solutions}

\subsection{Stellar Interior: Finch-Skea Solution}

We now solve the field equations \eqref{15}-\eqref{17} in pursuit of
spherical anisotropic solutions, by considering the sub field
equations \eqref{24}-\eqref{26} and \eqref{30}-\eqref{32}. A
solution of the general field equations \eqref{15}-\eqref{17} is
thus obtained by a linear combination of the solutions of the
aforementioned sub field equations, as suggested by the effective
parameters given by \eqref{19}. We begin with the system
\eqref{24}-\eqref{26}, for which we employ the Finch-Skea ansatz
\cite{43}
\begin{eqnarray}\label{41}
e^{\eta(r)}&=&\bigg[A+\frac{1}{2}\,B\,r\,\sqrt{Cr^2}\bigg]^2\,,
\\\label{42}
\sigma(r)&=&\frac{1}{1+Cr^2}\,,\\\nonumber \rho&=&\frac{12BC\lambda
r\left(Cr^2+1\right)\ln\left(A+\frac{1}{2}Br\sqrt{Cr^2}\right)}{\left(C
r^2+1\right)^2\left(2 A \sqrt{C r^2}+B C r^3\right)}\\\label{43}
&-&\frac{C\bigg(6\lambda-3+(4\lambda-1)Cr^2\bigg)\left(2A\sqrt{Cr^2}
+BCr^3\right)}{\left(Cr^2+1\right)^2\left(2A\sqrt{Cr^2}+BCr^3\right)},
\\\nonumber
P&=&\frac{C\bigg(6\lambda-1+(4\lambda-1)Cr^2\bigg)
\left(2A\sqrt{Cr^2}+BCr^3\right)}{\sqrt{Cr^2}\left(Cr^2+1\right)^2
\left(2A+Br\sqrt{Cr^2}\right)}\\\label{44}
&-&\frac{4BCr(3\lambda-1)\left(Cr^2+1\right)\ln\left(A+\frac{1}{2}
Br\sqrt{C r^2}\right)}{\sqrt{C r^2}\left(C r^2+1\right)^2\left(2A+Br
\sqrt{C r^2}\right)},
\end{eqnarray}
where the constants $A,~B$ and $C$ can be determined from the
matching conditions. This solution has been adopted because it is
both singularity-free as well as physically plausible. Choosing the
Schwarzschild metric as our exterior spacetime (i.e., for
$b^\ast(r)\rightarrow 0$ in Eq.\eqref{40}), the matching conditions
yield
\begin{eqnarray}\nonumber
A&=&\sqrt{\frac{R-2M_0}{R}}-\frac{R}{2}\,\sqrt{\frac{2M_0^2}{2R^3(R-2M_0)}}\,,
\\\label{51}
B&=&\sqrt{\frac{M_0}{2R^3}}\,, \quad C=\frac{1}{R^2-2R\,M_0}-\frac{1}{R^2}\,,
\end{eqnarray}
with the compactness ~$\frac{M_0}{2R}<\frac{2}{9}$. These values
ensure the surface continuity of the interior and exterior
geometries and will most certainly be altered upon addition of the
source $\Theta_{\tau\upsilon}$. We now find anisotropic solutions,
for which we shall set ($\delta\neq 0$) in the interior geometry and
utilize Eqs.\eqref{41} and \eqref{42} as our temporal and radial
metric coefficients, respectively. The deformation function
$h^\ast(r)$ is related to the source $\Theta_{\tau\upsilon}$ through
equations \eqref{30} to \eqref{32} which is a system of three
equations in four unknowns. Thus to close this system, we shall
impose a single constraint. We describe how to generate from the
physically acceptable isotropic Finch-Skea ansatz, new families of
anisotropic spherical solutions whose physical features are
inherited from the isotropic parent. We make it a point to mention
here that recently, many researchers have taken an interest in
exploring the gravitational decoupling scheme to extend known
isotropic solutions of self-gravitating systems to obtain
anisotropic spherical solutions in general relativity \cite{16,1a}
and vast modified theories \cite{1aa}-\cite{1bbb} including the
Rastall theory \cite{1c,1d}, as well as to obtain extended black
holes solutions \cite{1e,1f}.

\subsection{Solution I}

We shall impose a constraint on $\Theta_1^1$ and obtain a solution
of the field equations \eqref{30}-\eqref{32} for $h^\ast$ and
$\Theta_{\tau\upsilon}$. The interior geometry is compatible with an
exterior spacetime given by the Schwarzschild metric whenever
$b^\ast(R)\rightarrow 0$ in Eq.\eqref{39}, leading to the relation
$P(R)_{-}\sim\delta\left(\Theta_1^1(R)\right)_{-}$. It thus suffices
to choose
\begin{equation}\label{46}
\Theta^1_1(r)=P(r),
\end{equation}
which upon exploiting Eqs.\eqref{29} and \eqref{31} gives
\begin{equation}\label{47}
h^\ast(r)=-\sigma(r)+\bigg(\frac{1}{r^2}-G_\lambda(r)\bigg)
\bigg(\frac{\eta^\prime}{r}+\frac{1}{r^2}\bigg)^{-1},
\end{equation}
where we have used
$G_\lambda(r)=\lambda\bigg[\frac{4}{r^2}-\frac{\sigma^\prime}{r}
-\sigma\bigg(\frac{4}{r^2}+\frac{3\eta^\prime}{r}\bigg)\bigg]$ to
denote the Rastall contribution. We obtain the resulting expression
for the deformation function after the necessary simplifications as
\begin{align}\nonumber
h^\ast(r)&=\frac{4BCr^3(3\lambda -1)\left(C r^2+1\right)\ln
\left(A+\frac{1}{2}Br\sqrt{C r^2}\right)}{\left(Cr^2+1\right)^2
\left(BCr^3\left(4\ln\left(A+\frac{1}{2}Br\sqrt{C
r^2}\right)+1\right)+2A\sqrt{Cr^2}\right)}\\\label{48}
&-\frac{Cr^2\bigg((4 \lambda -1) Cr^2+6 \lambda -1\bigg) \left(2 A
\sqrt{Cr^2}+BC r^3\right)}{\left(C r^2+1\right)^2 \left(BC r^3
\left(4 \ln \left(A+\frac{1}{2} B r \sqrt{C r^2}\right)+1\right)+2 A
\sqrt{C r^2}\right)}\,.
\end{align}
The deformed radial metric component \eqref{23} can thus be
expressed as
\begin{equation}\label{49}
e^{-\beta}=(1-\delta)\sigma(r)+\delta\bigg(\frac{1}{r^2}
-G_\lambda(r)\bigg)\bigg(\frac{\eta^\prime}{r}+\frac{1}{r^2}
\bigg)^{-1}\,,
\end{equation}
which simplifies to
\begin{align}\nonumber
e^{-\beta}&=\frac{4 B C r^3 \left(C r^2+1\right) (\delta  (3 \lambda
-1)+1) \ln\left(A+\frac{1}{2} B r \sqrt{c r^2}\right)}{\left(C
r^2+1\right)^2 \left(B C r^3 \left(4 \ln \left(A+\frac{1}{2} B r
\sqrt{C r^2}\right)+1\right)+2 A \sqrt{C r^2}\right)}\\\label{50}
&-\frac{\left(2 A \sqrt{C r^2}+B C r^3\right) \left(C r^2
\left(\delta  \left((4 \lambda -1)Cr^2+6 \lambda
-1\right)-1\right)-1\right)}{\left(C r^2+1\right)^2 \left(B C r^3
\left(4 \ln\left(A+\frac{1}{2} B r \sqrt{C r^2}\right)+1\right)+2 A
\sqrt{C r^2}\right)}.
\end{align}
\begin{figure}\center
\epsfig{file=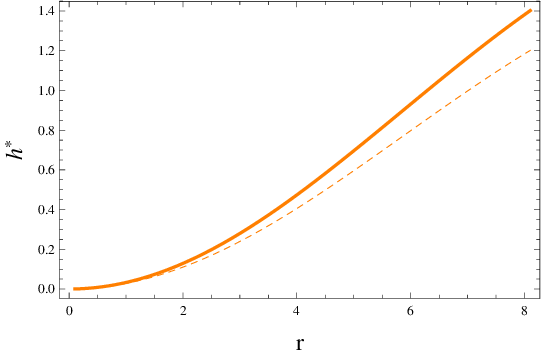,width=0.475\linewidth}
\epsfig{file=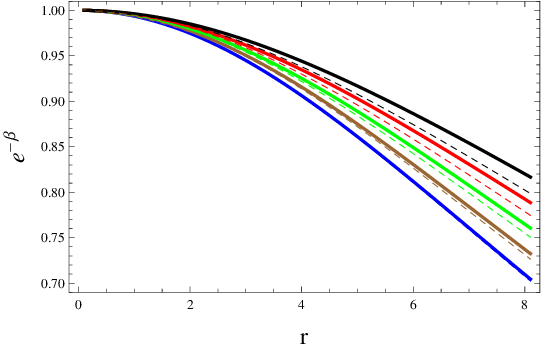,width=0.475\linewidth}\caption{Plots of $h^\ast$
and $e^{-\beta}$ versus $r$ corresponding to $\lambda=-0.4$ (solid),
$-0.5$ (dashed), $\delta=0.01$ (blue), $0.03$ (brown), $0.05$
(green), $0.07$ (red) and $0.09$ (black) for solution I.}
\end{figure}
\begin{figure}\center
\epsfig{file=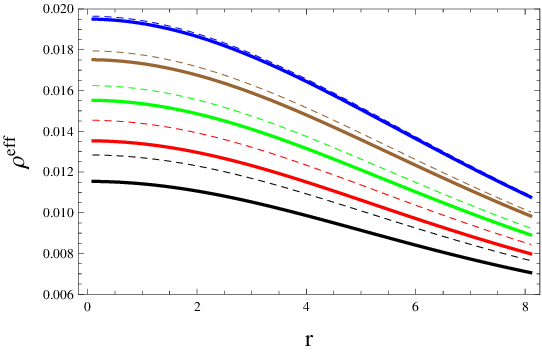,width=0.475\linewidth}
\epsfig{file=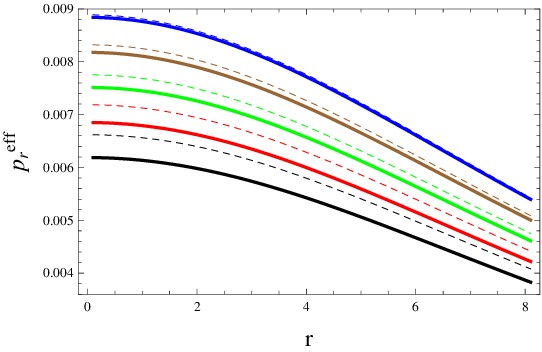,width=0.475\linewidth}
\epsfig{file=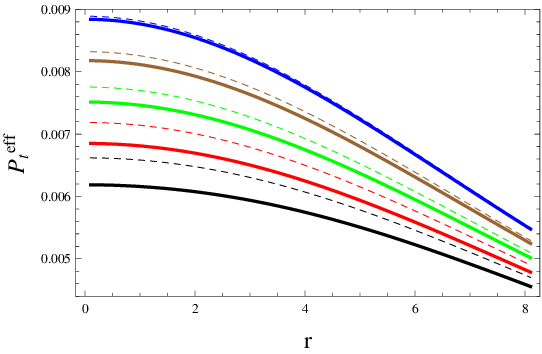,width=0.475\linewidth}
\epsfig{file=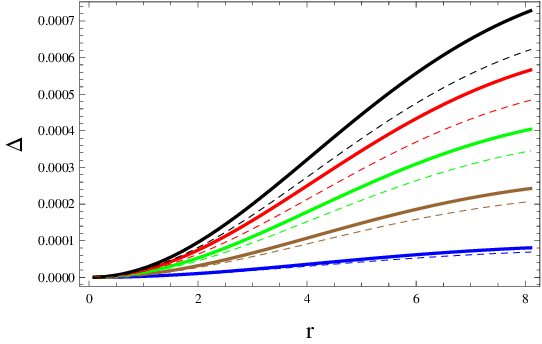,width=0.475\linewidth} \caption{Plots of
$\rho^{eff},P_r^{eff},P_t^{eff}$ and $\Delta$ versus $r$
corresponding to $\lambda=-0.4$ (solid), $-0.5$ (dashed),
$\delta=0.01$ (blue), $0.03$ (brown), $0.05$ (green), $0.07$ (red)
and $0.09$ (black) for solution I.}
\end{figure}
The interior metric functions \eqref{41} and \eqref{50} denote the
minimally deformed Finch-Skea solution by virtue of Eq.\eqref{48}.
We can now obtain the expressions for the effective parameters
together with the induced anisotropy, jointly constituting the
anisotropic solution. Due to lengthy expressions, we have displayed
these parameters in the appendix.

We now discuss the graphical analysis of the effective parameters
$\rho^{eff}$, $P_r^{eff}$, $P_t^{eff}$ and the anisotropy $\Delta$
for solution I. The graphical analysis is carried out using the star
candidate Her X-1 with mass $M_0=0.85M_{\bigodot}$ and radius
$R=8.1km$ \cite{44}. We use two values of the Rastall parameter,
given by $\lambda=-0.4,-0.5,$ and the coupling constant as
$\delta=0.01,0.03,0.05,0.07,0.09$. We mention here that these
parametric values are chosen after a long trial of values for which
they were found to induce the desired behavior in the graphical
analysis of the obtained models. We highlight the importance of
investigating the effect of the fluctuation of the Rastall and
decoupling parameters. It is essential to study the impact of the
Rastall parameter as it creates the sole deviation of Rastall theory
from general relativity. As for the decoupling parameter $\delta$,
its essence lies in the fact that it sets the stage for the
gravitational decoupling process, as it is through this parameter
that the anisotropic extra source is gravitationally coupled to the
isotropic seed source. Figure \textbf{1} shows the deformation
function $(h^\ast)$ and deformed radial coefficient. As expected
from the deformation function, it vanishes at the core. The behavior
of the effective parameters (energy density, radial and tangential
pressures) ought to be finite, positive and maximum at the center
whilst exhibiting a monotonically decreasing behavior towards the
star's surface, as shown in Figure \textbf{2}. It is observed that
the density attains lower values for the Rastall parameter
$\lambda=-0.4$ as compared to its other value, leading to the
conclusion that an increment in the Rastall parameter makes the
interior of compact star less dense. A similar observation is made
with regards to the radial and tangential pressures. In addition,
the radial and tangential pressures attain the same values at the
core and thus induce an anisotropy that vanishes at that point and
increases towards the surface. This positive anisotropy depicts an
outward directed pressure by virtue of which the anti-gravitational
force is produced, helping in stabilizing the compact structure. A
higher anisotropy is obtained with a reduction of the Rastall
parameter.

\subsection{Solution II}

Here we adopt a new constraint to derive a second anisotropic
solution. This constraint is imposed on the density parameter and is
taken to be
\begin{equation}\label{51}
\Theta_0^0(r)=\rho(r)\,.
\end{equation}
Using Eqs.\eqref{28} and \eqref{30} in the constraint above, we have
\begin{figure}\center
\epsfig{file=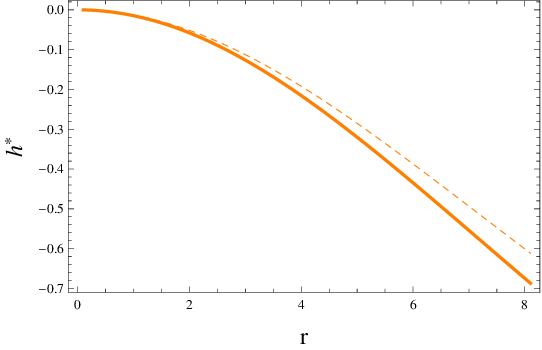,width=0.475\linewidth}
\epsfig{file=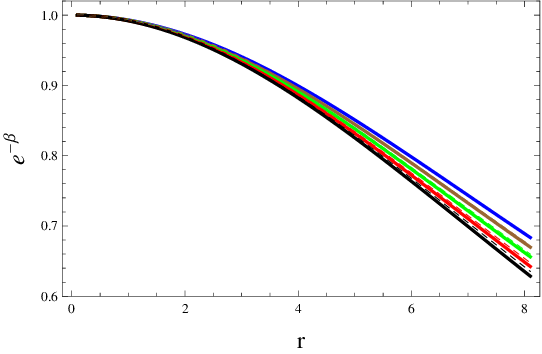,width=0.475\linewidth}\caption{Plots of $h^\ast$
and $e^{-\beta}$ versus $r$ corresponding to $\lambda=-0.4$ (solid),
$-0.5$ (dashed), $\delta=0.01$ (blue), $0.03$ (brown), $0.05$
(green), $0.07$ (red) and $0.09$ (black) for solution II.}
\end{figure}
\begin{figure}\center
\epsfig{file=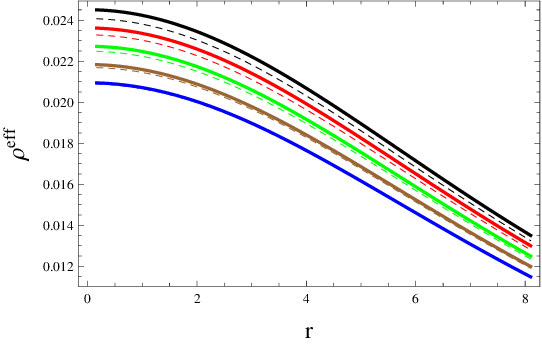,width=0.475\linewidth}
\epsfig{file=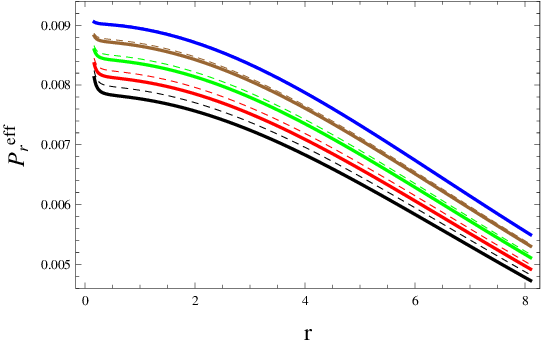,width=0.475\linewidth}
\epsfig{file=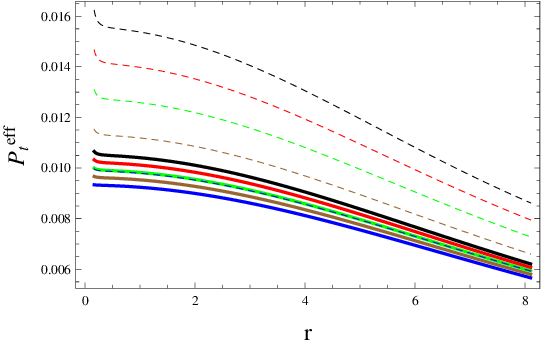,width=0.475\linewidth}
\epsfig{file=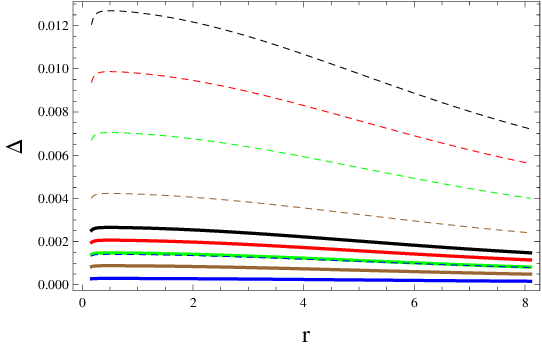,width=0.475\linewidth} \caption{Plots of
$\rho^{eff},P_r^{eff},P_t^{eff}$ and $\Delta$ versus $r$
corresponding to $\lambda=-0.4$ (solid), $-0.5$ (dashed),
$\delta=0.01$ (blue), $0.03$ (brown), $0.05$ (green), $0.07$ (red)
and $0.09$ (black) for solution II.}
\end{figure}
\begin{equation}\label{52}
-\frac{h^{\ast^\prime}}{r}-\frac{h^\ast}{r^2}=
\frac{1}{r^2}-\frac{\sigma^\prime}{r}-\frac{\sigma}{r^2}
-\lambda\bigg[\frac{4}{r^2}-\frac{\sigma^\prime}{r}
-\sigma\bigg(\frac{3\eta^\prime}{r}+\frac{4}{r^2}\bigg)\bigg]\,.
\end{equation}
Due to the unavailability of an exact solution of the differential
equation \eqref{52}, we proceed with a numerical approximation. The
graphical description for solution II follows exactly as in the case
of solution I. The deformation function and the deformed radial
coefficient are obtained and shown in Figure \textbf{3}. The
effective parameters and anisotropy are also obtained and plotted in
Figure \textbf{4}. As the case of the solution I, they exhibit a
behavior consistent with compact stars (i.e., positive, finite,
maximum at the core and monotonically decreasing towards the
boundary). However, in contrast to solution I, it is observed here
that an increment in the Rastall parameter makes the interior of
compact stars more dense. The effect of the increment in the Rastall
parameter is seen to coincide with a reduction in the radial and
tangential pressures. Due to the inequality of the radial and
tangential pressures at the core, the corresponding anisotropy is
non-vanishing at the core and possesses a positive profile
everywhere.

\subsection{Analysis of Physical Viability and Stability}

Here, we shall investigate various physical features of both
solutions, ranging from physical viability to stability. The energy
conditions are physical restrictions imposed on the stress-energy
tensor and (if satisfied) portray the existence of ordinary matter
in the interior of stellar distribution. These conditions can be
classified as dominant, strong, weak and null energy conditions as
follows.
\begin{itemize}
\item Dominant Energy Conditions\\
$\rho^{eff}-P_r^{eff}\geq 0,\quad\rho^{eff}-P_t^{eff}\geq 0.$

\item Strong Energy Conditions\\
$\rho^{eff}+P_r^{eff}\geq 0,\quad\rho^{eff}+P_t^{eff}\geq
0,\quad\rho^{eff}+P_r^{eff}+2P_t^{eff}\geq 0.$

\item Weak Energy Conditions\\
$\rho^{eff}\geq 0,\quad\rho^{eff}+P_r^{eff}\geq
0,\quad\rho^{eff}+P_t^{eff}\geq 0.$

\item Null Energy Conditions\\
$\rho^{eff}+P_r^{eff}\geq 0,\quad\rho^{eff}+P_t^{eff}\geq 0.$
\end{itemize}
As seen in Figures \textbf{5} and \textbf{6}, all the energy bounds
are met by our obtained solutions thereby implying physical
viability. We also investigate an interesting physical feature of
celestial objects, dubbed the EoS parameter. With respect to the
radial and tangential EoS denoted by
$\omega_r=\frac{P_r^{eff}}{\rho^{eff}}$ and
$\omega_t=\frac{P_t^{eff}}{\rho^{eff}}$, respectively, an effective
stellar configuration is implied if $0\leq\omega_r\leq 1$ and
$0\leq\omega_t\leq 1$ \cite{45}. This condition is also satisfied by
both solutions as shown in Figure \textbf{7}.
\begin{figure}\center
\epsfig{file=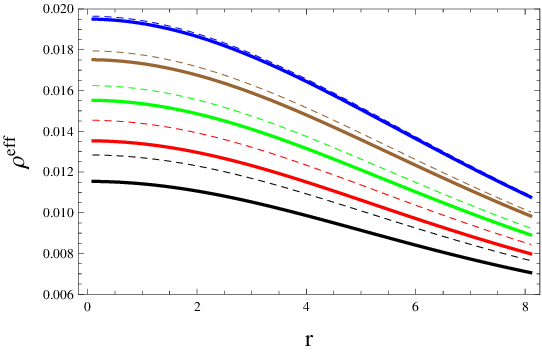,width=0.475\linewidth}
\epsfig{file=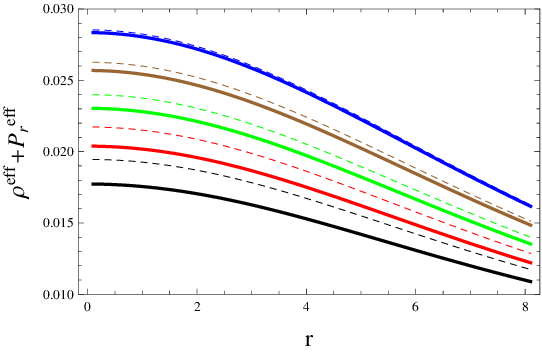,width=0.475\linewidth}
\epsfig{file=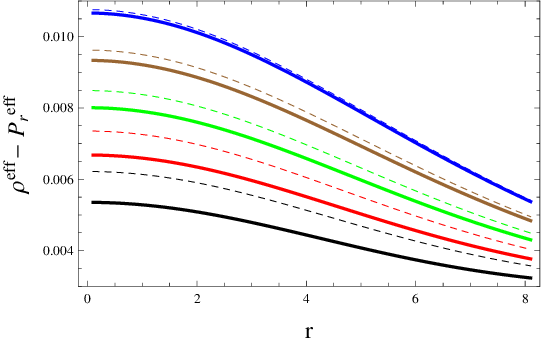,width=0.475\linewidth}
\epsfig{file=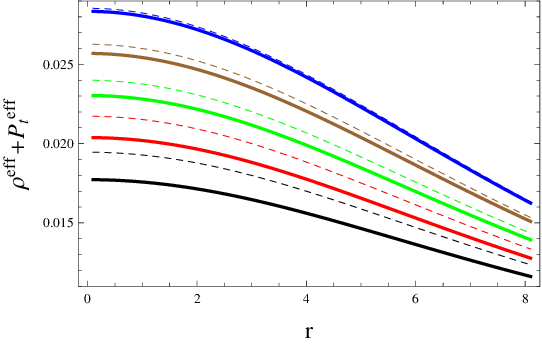,width=0.475\linewidth}
\epsfig{file=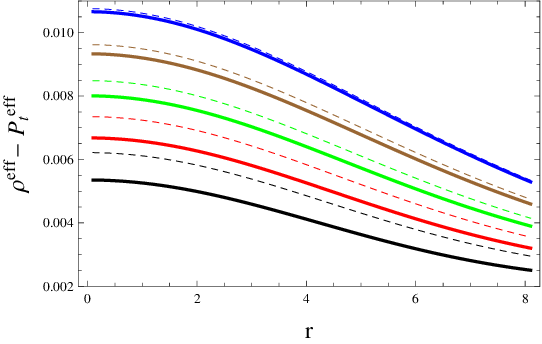,width=0.475\linewidth}
\epsfig{file=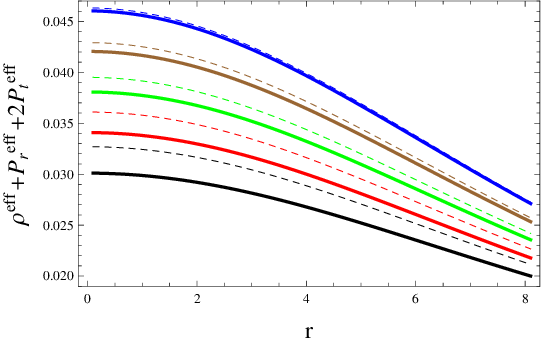,width=0.475\linewidth} \caption{Plots of energy
conditions versus $r$ corresponding to $\lambda=-0.4$ (solid),
$-0.5$ (dashed), $\delta=0.01$ (blue), $0.03$ (brown), $0.05$
(green), $0.07$ (red) and $0.09$ (black) for solution I.}
\end{figure}
\begin{figure}\center
\epsfig{file=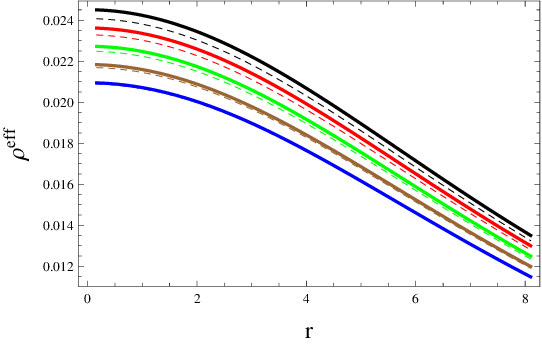,width=0.475\linewidth}
\epsfig{file=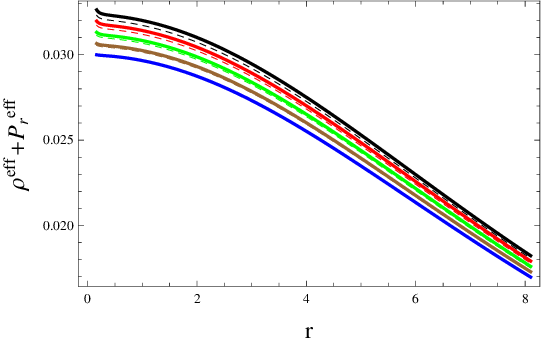,width=0.475\linewidth}
\epsfig{file=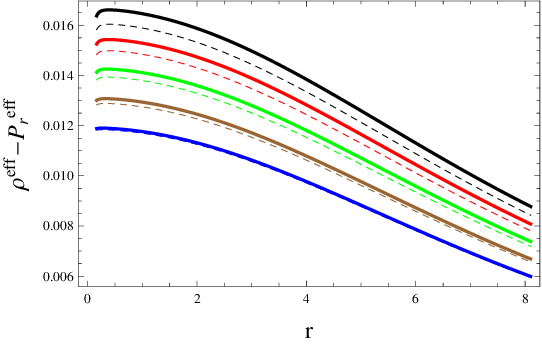,width=0.475\linewidth}
\epsfig{file=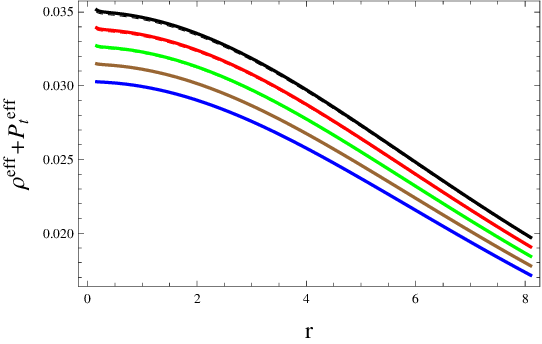,width=0.475\linewidth}
\epsfig{file=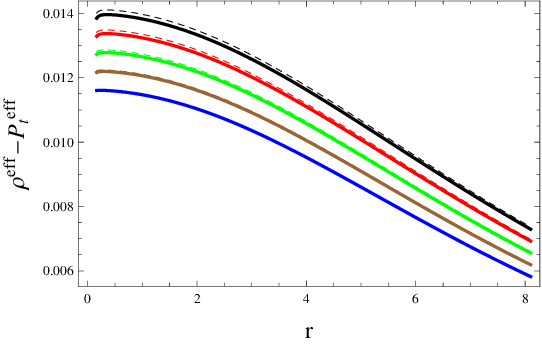,width=0.475\linewidth}
\epsfig{file=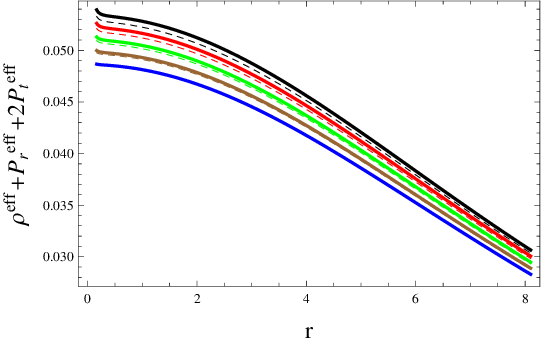,width=0.475\linewidth} \caption{Plots of energy
conditions versus $r$ corresponding to $\lambda=-0.4$ (solid),
$-0.5$ (dashed), $\delta=0.01$ (blue), $0.03$ (brown), $0.05$
(green), $0.07$ (red) and $0.09$ (black) for solution II.}
\end{figure}

Additionally, we examine the compactness $u(r)=\frac{m(r)}{r}$ and
surface redshift $Z_s=\frac{1}{\sqrt{1-2u(r)}}-1$. The compactness
of a celestial body describes how densely its mass is packed within
a specific volume or radius. This measure is dimensionless and helps
to determine the intensity of the gravitational field at the
object's surface. On the other hand, surface redshift refers to the
shift in the wavelength of light or electromagnetic radiation
emitted from the surface of a dense object when viewed from a
distance. The powerful gravitational pull near the object's surface
causes the light to lose energy, resulting in a longer wavelength,
or redshift. With these parameters, the limits $u(r)<\frac{4}{9}$
\cite{45a}~and~$Z_s\leq 5.2$ \cite{46} guaranty an effective matter
configuration. It can be observed that the surface redshift is
dependent on the compactness function which, in turn, depends on the
mass function. The mass of the sphere can be determined by the
equation
\begin{equation}\label{53}
m(r)=4\pi\int_0^r \rho^{eff}r^2 dr.
\end{equation}
Both solutions satisfy the stated compactness and surface redshift
limits as shown in their plots displayed in Figure \textbf{8}.
\begin{figure}\center
\epsfig{file=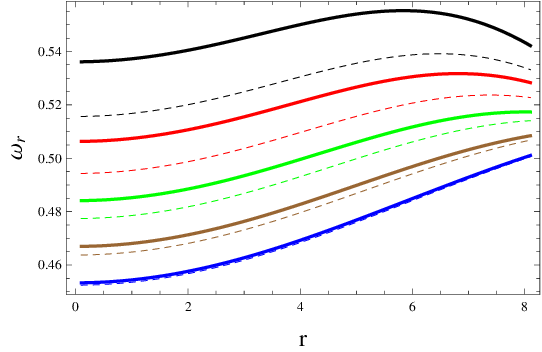,width=0.475\linewidth}
\epsfig{file=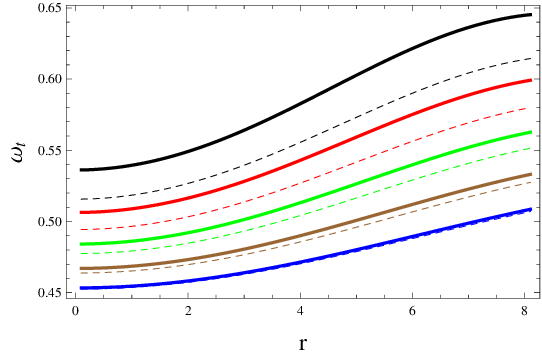,width=0.475\linewidth}
\epsfig{file=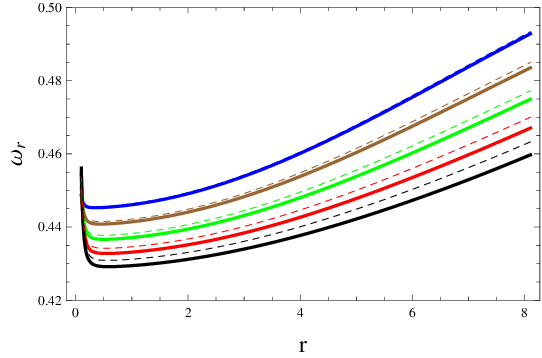,width=0.475\linewidth}
\epsfig{file=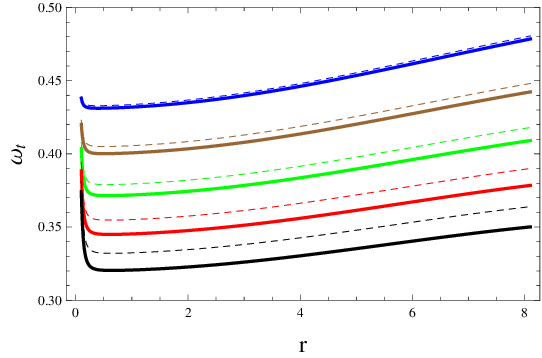,width=0.475\linewidth}\caption{Plots of
radial and tangential EoS parameters versus $r$ corresponding to
$\lambda=-0.4$ (solid), $-0.5$ (dashed), $\delta=0.01$ (blue),
$0.03$ (brown), $0.05$ (green), $0.07$ (red) and $0.09$ (black) for
solutions I (top row) and II (bottom row).}
\end{figure}
\begin{figure}\center
\epsfig{file=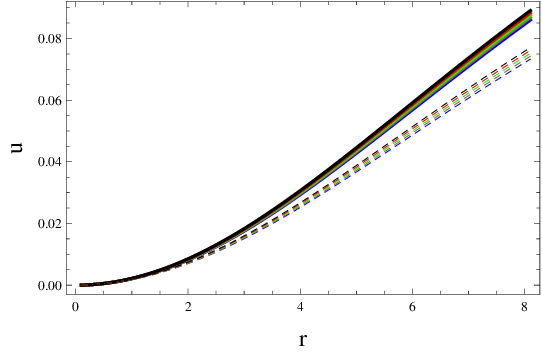,width=0.475\linewidth}
\epsfig{file=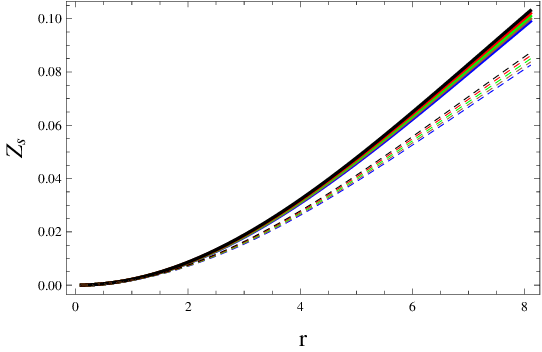,width=0.475\linewidth}
\epsfig{file=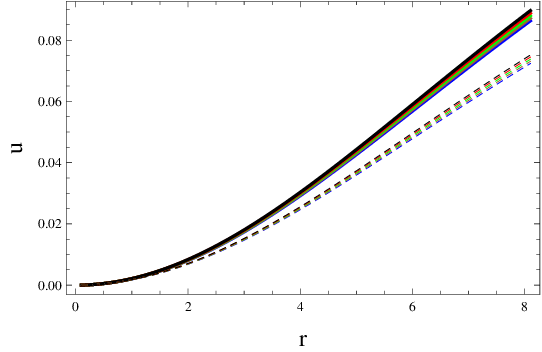,width=0.475\linewidth}
\epsfig{file=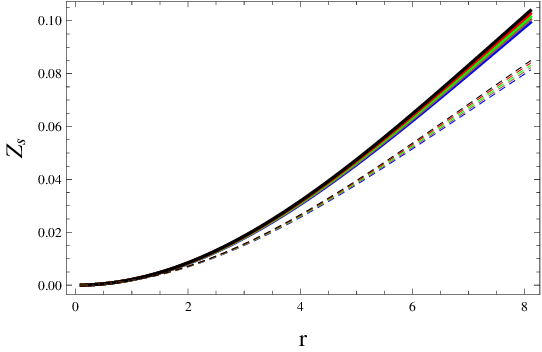,width=0.475\linewidth}\caption{Plots of
compactness and surface redshift versus $r$ corresponding to
$\lambda=-0.4$ (solid), $-0.5$ (dashed), $\delta=0.01$ (blue),
$0.03$ (brown), $0.05$ (green), $0.07$ (red) and $0.09$ (black) for
solutions I (top row) and II (bottom row).}
\end{figure}
\begin{figure}\center
\epsfig{file=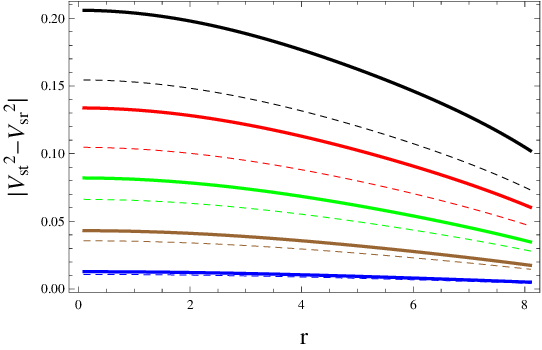,width=0.475\linewidth}
\epsfig{file=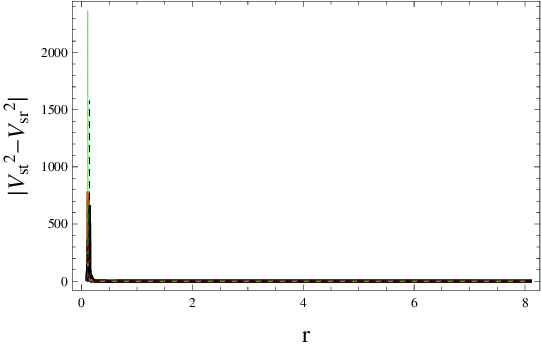,width=0.475\linewidth}\caption{Plots of
$|V_{st}^2-V_{sr}^2|$ versus $r$ corresponding to $\lambda=-0.4$
(solid), $-0.5$ (dashed), $\delta=0.01$ (blue), $0.03$ (brown),
$0.05$ (green), $0.07$ (red) and $0.09$ (black) for solutions I
(left) and II (right).}
\end{figure}

We now shift our focus to the stability analysis of the obtained
solutions. We first use the Herrera cracking technique \cite{20} in
which stability demands that $0\leq |V_{st}^2-V_{sr}^2|\leq 1$,
where $V_{st}^2=\frac{dP_t^{eff}}{d\rho^{eff}}$ and
$V_{sr}^2=\frac{dP_r^{eff}}{d\rho^{eff}}$ denote the tangential and
radial sound speeds, respectively. Through this test, we show that
solution I is stable while solution II is unstable (Figure
\textbf{9}). The stability of both solutions is further tested using
the causality condition wherein stability necessitates that the
speed of sound components must be contained in the range
$\left[0,~1\right]$, i.e., $0\leq V_{sr}^2\leq 1$ and $0\leq
V_{st}^2\leq 1\,.$ The results of this test which are shown in
Figure \textbf{10}, corroborates the outcome of the Herrera cracking
test. The plot of the Herrera cracking condition for solution II (in
the right panel of Figure \textbf{9}) as well as the plots of the
causality conditions for solution II (in the bottom panel of Figure
\textbf{10}) display unbounded behavior at the core. This behavior
at the core can be attributed to the instability of the said model.
We highlight that the discontinuity portrayed at the core, in some
of the matter variables (for solution II) plotted in Figures
\textbf{4} and \textbf{6} are in line with the obtained results for
the stability of this model. This behavior is the characteristic of
an unstable model as deduced through the aforementioned stability
analysis.
\begin{figure}\center
\epsfig{file=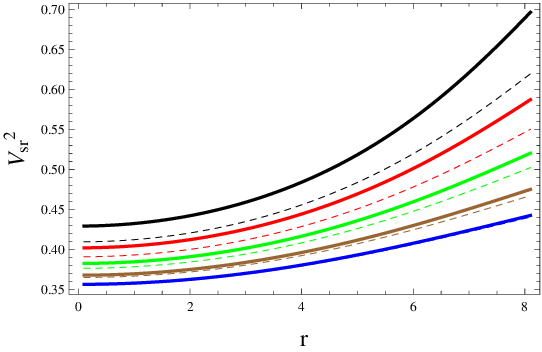,width=0.475\linewidth}
\epsfig{file=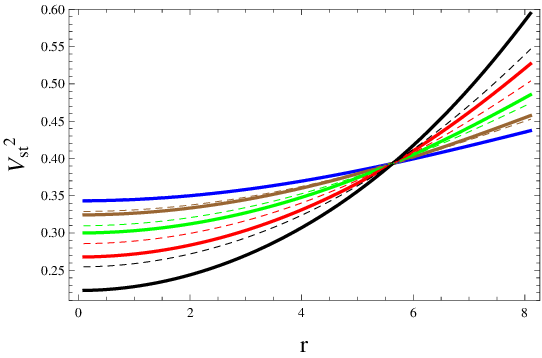,width=0.475\linewidth}
\epsfig{file=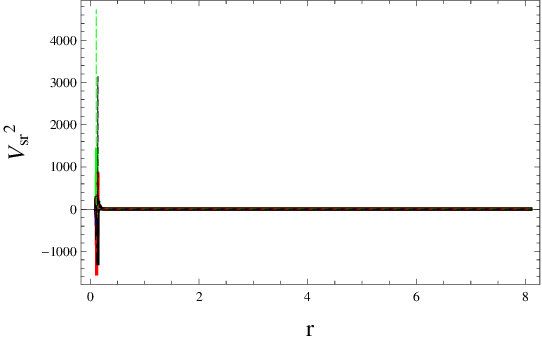,width=0.475\linewidth}
\epsfig{file=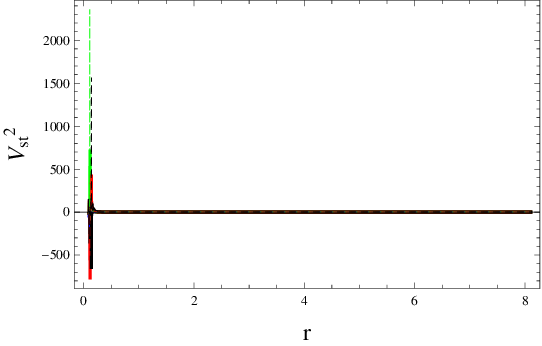,width=0.475\linewidth}\caption{Plots of radial and
tangential sound speeds versus $r$ corresponding to $\lambda=-0.4$
(solid), $-0.5$ (dashed), $\delta=0.01$ (blue), $0.03$ (brown),
$0.05$ (green), $0.07$ (red) and $0.09$ (black) for solutions I (top
row) and II (bottom row).}
\end{figure}

\section{Conclusions}

Numerous researchers are involved in the search for interior
solutions defining self-gravitating systems. To this end,
astrophysicists have made several efforts to build stable and
physically viable solutions for compact objects. Recently, the MGD
technique has been widely applied to obtain precise solutions for
the internal constitution of stellar objects. In this paper, we have
used this approach to obtain anisotropic spherical solutions by
extending a known isotropic interior solution, namely, the
Finch-Skea ansatz. The static and spherically symmetric Rastall
field equations \eqref{15}-\eqref{17} have been decoupled into two
sets, the first set corresponding to the Rastall field equations
\eqref{24}-\eqref{26} for isotropic matter distribution
$\bar{T}_{\tau\upsilon}$ while the second set \eqref{30}-\eqref{32}
characterizes the anisotropic source $\Theta_{\tau\upsilon}\,.$ The
junction conditions that govern the smooth matching at the stellar
surface have also been studied, taking the exterior geometry to be
the Schwarzschild spacetime.

Since the Rastall theory of gravity contains an extra term that
deviates from general relativity, we have investigated the effects
of this term in this work. To extend the Finch-Skea solution to an
anisotropic domain, we have followed the same procedure given in
\cite{16}. Using the mimic constraint approach, we impose suitable
conditions that relate the thermodynamic seed variables with the
corresponding components of the $\Theta$-sector so that the
decoupling function $h^\ast(r)$ can be determined. We have thus used
two constraints: a pressure-like constraint in which the $1-1$
component of the $\Theta$-sector mimics the seed pressure $P(r)$,
and a density-like constraint in which the $0-0$ component mimics
seed energy density $\rho(r)$. In the case of solution I, a simple
algebraic equation has been obtained from which an explicit
expression for the decoupling function is easily derived. However,
for solution II, a first order differential equation has appeared
from which a numerical solution of the decoupling function is
obtained due to mathematical complications introduced by the Rastall
contribution $G_\lambda(r)$.

For the Rastall parameter $\lambda=-0.4,-0.5~$ with the decoupling
constant $~\delta=0.01,0.03,0.05,0.07,0.09,$ the physical behavior
of the effective parameters for both solutions have been found to be
in agreement with the requirements for compact stars. The generated
anisotropy in both cases has been found to be positive, implying an
outward directed pressure that produces the anti-gravitational force
necessary to keep the compact object in an equilibrium state. We
have found that increasing the value of $\lambda$ provides a less
dense interior of compact stars in the case of solution I and a more
dense interior corresponding to solution II. We have also found that
increasing the value of the decoupling parameter $\delta$ enhances a
less dense interior of compact stars in the case of solution I and a
more dense interior corresponding to solution II. In addition, the
physical viability of both solutions has been endorsed through
analysis of the energy conditions. Stability analysis has also been
done through the Herrera cracking and causality condition through
which we have established that solution I is stable while solution
II is unstable. We would like to point out that such tests for
physical viability and stability have not been executed in the case
of general relativity \cite{16}. Finally, all our results can be
reduced to general relativity for $\lambda=0$.
\section*{Appendix: Effective parameters $\rho^{eff},P_r^{eff},P_t^{eff}$
and Anisotropy $\Delta$}
\renewcommand{\theequation}{A\arabic{equation}}
\setcounter{equation}{0}
\begin{align}\nonumber
\rho^{eff}&=\frac{C\sqrt{Cr^2}}{\left(C r^2+1\right)^3 \left(2 A+B r
\sqrt{C r^2}\right)}\\\nonumber &\times\bigg[\frac{\left(C
r^2+1\right) \left(C r^2+3\right) \left(4 A \left(A+B r \sqrt{C
r^2}\right)+B^2 C r^4\right)}{2 A \sqrt{C r^2}+B C r^3}\\\nonumber
&+\frac{C \sqrt{C r^2}}{\left(B C r^3 \left(4 \ln
\left(A+\frac{1}{2} B r \sqrt{C r^2}\right)+1\right)+2 A
\sqrt{Cr^2}\right)^2}\\\nonumber &\times\bigg(8 A^3 \sqrt{C r^2}
\left(C r^2 \left(C (4 \lambda -1) r^2+14 \lambda -4\right)+18
\lambda -3\right)\\\nonumber &+4 B r \ln\left(A+\frac{1}{2} B r
\sqrt{C r^2}\right)\bigg(4 A^2 \bigg(C r^2 \bigg(C r^2 \left(C(1-4
\lambda ) r^2-9 \lambda +1\right)\\\nonumber &-6 \lambda +3\bigg)-9
\lambda +3\bigg)+4 Br (3 \lambda -1) \left(C^2 r^4-1\right)\left(2 A
\sqrt{C r^2}+B C r^3\right)\\\nonumber &\times
\ln\left(A+\frac{1}{2} B r \sqrt{C r^2}\right)+8 A B r \sqrt{C r^2}
\bigg(C r^2 \left(C (2 \lambda -1) r^2+3 \lambda \right)\\\nonumber
&-3 \lambda +1\bigg)+ B^2 C r^4 \bigg(C r^2 \left(C r^2 \left(C (4
\lambda -1) r^2+17 \lambda -5\right)+18 \lambda -3\right)\\\nonumber
&-3 \lambda +1\bigg)\bigg)+12 A^2 B C r^3 \left(C r^2 \left(C (4
\lambda -1) r^2+14 \lambda -4\right)+18 \lambda -3\right)\\\nonumber
&+2 A B^2 r^2 \sqrt{C r^2} \bigg(C r^2 \left(C r^2 \left(5 C (1-4
\lambda) r^2-62 \lambda +12\right)-42 \lambda +15\right)\\\nonumber
&-24 \lambda +8\bigg)+B^3 C r^5 \bigg(C r^2 \bigg(C r^2 \left(7
C(1-4 \lambda) r^2-90 \lambda +20\right)\\\nonumber &-78 \lambda
+21\bigg)-24 \lambda +8\bigg)\bigg)\bigg],
\end{align}
\begin{align}\nonumber
P_r^{eff}&=\frac{4 BCr \left(Cr^2+1\right) (\delta  (3 \lambda
-1)+1) \ln \left(A+\frac{1}{2} B r \sqrt{Cr^2}\right)}{\sqrt{Cr^2}
\left(Cr^2+1\right)^2 \left(2 A+B r \sqrt{Cr^2}\right)}\\\nonumber
&-\frac{C\left(2 A \sqrt{Cr^2}+B cr^3\right) \bigg(C r^2 (4 \delta
\lambda -\delta +1)+6 \delta \lambda -\delta +1\bigg)}{\sqrt{C r^2}
\left(C r^2+1\right)^2 \left(2 A+B r \sqrt{C r^2}\right)},
\end{align}
\begin{align}\nonumber
P_t^{eff}&=\frac{C}{\left(C r^2+1\right)^3 \left(2 A+B r \sqrt{C
r^2}\right)^2}\\\nonumber &\times\bigg[\bigg(\frac{(1+Cr^2)^24 B
\left(2 A \sqrt{C r^2}+B C r^3\right) \ln \left(A+\frac{1}{2} B r
\sqrt{C r^2}\right)}{Cr}\\\nonumber &-\frac{(1+Cr^2)^2C r \left(4 A
\left(A+B r \sqrt{C r^2}\right)+B^2
Cr^4\right)}{Cr}\bigg)\\\nonumber &+\frac{\delta}{\left(B C r^3
\left(4 \ln \left(A+\frac{1}{2} B r \sqrt{C r^2}\right)+1\right)+2 A
\sqrt{C r^2}\right)^2}\\\nonumber &\times \bigg(\left(B C r^3
\left(2 \ln \left(A+\frac{1}{2} B r \sqrt{C r^2}\right)+1\right)+2 A
\sqrt{C r^2}\right)\\\nonumber &\times \bigg(8 A^3 \sqrt{C r^2}
\left(C (1-2 \lambda ) r^2-6 \lambda +1\right)+4 B r \ln
\left(A+\frac{1}{2} B r \sqrt{C r^2}\right)\\\nonumber &\times
\bigg(4 A^2 \left(C r^2 \left(C r^2 \left(C (4 \lambda -1) r^2+8
\lambda -1\right)+3 \lambda -1\right)+3 \lambda -1\right)\\\nonumber
&+2 A B r \sqrt{C r^2} \left(C r^2 \left(C r^2 \left(C (4 \lambda
-1) r^2+3 \lambda +1\right)-6 \lambda +1\right)+3 \lambda
-1\right)\\\nonumber &-4 B C (3 \lambda -1) r^3 \left(C r^2+1\right)
\left(2 A \sqrt{C r^2}+B C r^3\right) \ln \left(A+\frac{1}{2} B r
\sqrt{C r^2}\right)\\\nonumber &+B^2 C^2 r^6 \left(C (2-5 \lambda )
r^2-9 \lambda +2\right)\bigg)+12 A^2 B C r^3 \left(C (1-2 \lambda )
r^2-6 \lambda +1\right)\\\nonumber &+2 A B^2 r^2 \sqrt{C r^2}
\bigg(C r^2 \left(C r^2 \left(4 C (4 \lambda -1) r^2+46 \lambda
-9\right)+30 \lambda -9\right)\\\nonumber &+12 \lambda -4\bigg)+B^3
C r^5 \bigg(C r^2 \left(C r^2 \left(4 C (4 \lambda -1) r^2+50
\lambda -11\right)+42 \lambda -11\right)\\\nonumber &+12 \lambda
-4\bigg)\bigg)-4 B r \sqrt{C r^2} \left(C r^2+1\right) \bigg(B r
\sqrt{C r^2} 2\ln\left(A+\frac{1}{2} B r \sqrt{C
r^2}\right)\\\nonumber &+2 A \ln \left(A+\frac{1}{2} B r \sqrt{C
r^2}\right)+B r \sqrt{C r^2}\bigg) \bigg(B C r^3 \bigg(4 \ln
\left(A+\frac{1}{2} B r \sqrt{C r^2}\right)\\\nonumber &+1\bigg)+2 A
\sqrt{C r^2}\bigg) \bigg(\left(C (4 \lambda -1) r^2+6 \lambda
-1\right) \left(2 A \sqrt{C r^2}+B C r^3\right)\\\nonumber &-4 B (3
\lambda -1) r \left(C r^2+1\right) \ln \left(A+\frac{1}{2} B r
\sqrt{C r^2}\right)\bigg)\bigg) \bigg],
\end{align}
\begin{align}\nonumber
\Delta&=\frac{\delta\,C}{\left(C r^2+1\right)^3 \left(2 A+B r
\sqrt{C r^2}\right)^2}\bigg[\bigg(\frac{\left(C r^2+1\right) \left(2
A+B r \sqrt{C r^2}\right)}{\sqrt{Cr^2}}\\\nonumber
&\times\bigg(\left(C (4 \lambda -1) r^2+6 \lambda -1\right) \left(2
A \sqrt{C r^2}+B C r^3\right)\\\nonumber &-4 B (3 \lambda -1) r
\left(C r^2+1\right) \ln \left(A+\frac{1}{2} B r \sqrt{C
r^2}\right)\bigg) \bigg)\\\nonumber &-\frac{1}{B C r^3 \left(4 \ln
\left(A+\frac{1}{2} B r \sqrt{C r^2}\right)+1\right)+2 A \sqrt{C
r^2}}\\\nonumber &\times4 B r \sqrt{C r^2} \left(C r^2+1\right)
\bigg(B r \sqrt{C r^2} \ln ^2\left(A+\frac{1}{2} B r \sqrt{C
r^2}\right)\\\nonumber &+2 A \ln \left(A+\frac{1}{2} B r \sqrt{C
r^2}\right)+B r \sqrt{C r^2}\bigg) \bigg(\left(C (4 \lambda -1)
r^2+6 \lambda -1\right)\\\nonumber &\times \left(2 A \sqrt{C r^2}+B
C r^3\right)-4 B (3 \lambda -1) r \left(C r^2+1\right)\ln
\left(A+\frac{1}{2} B r \sqrt{C r^2}\right)\bigg)\\\nonumber
&+\frac{1}{\left(B C r^3 \left(4 \ln \left(A+\frac{1}{2} B r \sqrt{C
r^2}\right)+1\right)+2 A \sqrt{C r^2}\right)^2}\bigg(B C r^3 \bigg(2
\ln \bigg(A\\\nonumber &+\frac{1}{2} B r \sqrt{C
r^2}\bigg)+1\bigg)+2 A \sqrt{C r^2}\bigg) \bigg(8 A^3 \sqrt{C r^2}
\left(C (1-2 \lambda ) r^2-6 \lambda +1\right)\\\nonumber &+4 B r
\ln \left(A+\frac{1}{2} B r \sqrt{C r^2}\right) \bigg(4 A^2 \bigg(C
r^2 \left(C r^2 \left(C (4 \lambda -1) r^2+8 \lambda -1\right)+3
\lambda -1\right)\\\nonumber &+3 \lambda -1\bigg)+2 A B r \sqrt{C
r^2} \left(c r^2 \left(C r^2 \left(C (4 \lambda -1) r^2+3 \lambda
+1\right)-6 \lambda +1\right)+3 \lambda -1\right)\\\nonumber &-4 B C
(3 \lambda -1) r^3 \left(C r^2+1\right) \left(2 A \sqrt{C r^2}+B C
r^3\right) \ln \left(A+\frac{1}{2} B r \sqrt{C r^2}\right)+B^2 C^2
r^6\\\nonumber &\times \left(C (2-5 \lambda ) r^2-9 \lambda
+2\right)\bigg)+12 A^2 B C r^3 \left(C (1-2 \lambda ) r^2-6 \lambda
+1\right)\\\nonumber &+2 A B^2 r^2 \sqrt{C r^2} \bigg(C r^2 \left(C
r^2 \left(4 C (4 \lambda -1) r^2+46 \lambda -9\right)+30 \lambda
-9\right)\\\nonumber &+12 \lambda -4\bigg)+B^3 C r^5 \bigg(C r^2
\bigg(C r^2 \bigg(4 C (4 \lambda -1) r^2+50 \lambda
-11\bigg)\\\nonumber &+42 \lambda -11\bigg)+12 \lambda
-4\bigg)\bigg) \bigg].
\end{align}\\\\
\textbf{Data Availability Statement:} No data was used for the
research described in this paper.

\end{document}